# Title: Survival Function Analysis of Planet Size Distribution

**Authors:** Li Zeng*[1], Stein B. Jacobsen[1], Dimitar D. Sasselov[2], Andrew Vanderburg[3]

**Affiliations:**

[1]Department of Earth & Planetary Sciences, Harvard University, 20 Oxford Street, Cambridge, MA 02138.

[2]Harvard-Smithsonian Center for Astrophysics, 60 Garden Street, Cambridge, MA 02138,

[3]Sagan Fellow, University of Texas at Austin

Correspondence to: astrozeng@gmail.com

**Abstract:**

Applying the survival function analysis to the planet radius distribution of the *Kepler* exoplanet candidates, we have identified two natural divisions of planet radius at 4 Earth radii ($R_\oplus$) and 10 $R_\oplus$. These divisions place constraints on planet formation and interior structure model. The division at 4 $R_\oplus$ separates small exoplanets from large exoplanets above. When combined with the recently-discovered radius gap at 2 $R_\oplus$, it supports the treatment of planets 2-4 $R_\oplus$ as a separate group, likely water worlds. Thus, for planets around solar-type FGK main-sequence stars, we argue that 2 $R_\oplus$ is the separation between water-poor and water-rich planets, and 4 $R_\oplus$ is the separation between gas-poor and gas-rich planets. We confirm that the slope of survival function in between 4 and 10 $R_\oplus$ to be shallower compared to either ends, indicating a relative paucity of planets in between 4-10 $R_\oplus$, namely, the sub-Saturnian desert there. We name them transitional planets, as they form a bridge between the gas-poor small planets and gas giants. Accordingly, we propose the following classification scheme: (<2 $R_\oplus$) rocky planets, (2-4 $R_\oplus$) water worlds, (4-10 $R_\oplus$) transitional planets, and (>10 $R_\oplus$) gas giants.

**Method:**

    **(1) Survival Function Analysis**

The survival function (Clauset et al., 2009; Feigelson and Nelson, 1985; Virkar and Clauset, 2014), also known as the complimentary cumulative distribution function (cCDF), is defined in this context as the number of planets above a given radius, versus radius in a log-log plot, of the *Kepler* planet candidates (4433 from Q1-Q17 DR 25 of *NASA* Exoplanet Archive (Akeson et al., 2013), versus 1861 from California-Kepler Survey (Fulton et al., 2017) with improved stellar parameters, both with false positives excluded already).

SF (Survival Function) = 1- CDF (Cumulative Distribution Function) = 1 - Integral of PDF (Probability Density Function)

Differentiate SF, one gets the PDF (Probability Density Function).

The survival function (SF) can tell apart different distributions. Comparing to the probability density function (PDF), it has the advantage of overcoming the large fluctuations that occur in the tail of a distribution due to finite sample sizes (Clauset et al., 2009). For example, on a log-log plot of survival function, power-law distribution appears as a straight line, while normal, log-normal, or exponential distributions all have a sharp cut-off (upper bound) in radius. This plot is also known as the rank-frequency plot (Newman, 2005). The comparison of the SF of the data with the straight-line SF of reference power-law distributions in **Figure 1** is the essence of the Kolmogorov-Smirnov (K-S) test. K-S statistic simply evaluates the maximum distance between the CDFs of the data and the fitted model and can also test the goodness-of-fit. This approach identifies the boundaries separating different regimes of distributions in the data.

### (2) Error Analysis

In **Figure 2**, we use a Monte-Carlo method to determine the uncertainty in the survival function. For example, each of the planet radii has some best-fit value, and some uncertainty (for example, $1 \pm 0.1~R_\oplus$, $2.3 \pm 0.05~R_\oplus$, etc). What we have done so far in **Figure 1** is to calculate the survival function with the best-fit values (in those two examples, we would use 1, and 2.3, for example). In **Figure 2**, for each planet we randomly draw a number from the asymmetric Gaussian distribution centered at the mean value with a width equal to the uncertainty in plus and minus direction in the radius measurement. For those two examples, we might randomly draw 1.03 and $2.34~R_\oplus$. Then, calculate the survival function with these newly drawn radius measurements. Then, repeat that whole process 100 times, and calculate 100 survival functions. This gives a good idea as to the uncertainty in the survival function itself.

In **Figure 2**, we also perform various cuts on the datasets (both KOI and CKS), with the same selection steps and criteria presented in (Fulton et al., 2017). The sequential cuts come at the expense of losing some potentially valuable data points and suffer more and more small number statistics and fluctuations towards larger radius. So, there is a trade-off. As shown in **Figure 2**, the breaks at $4~R_\oplus$ and $10~R_\oplus$ are robust. And the identification of the slope of about -1 in the survival function of planets in between $4$-$10~R_\oplus$ is also robust.

In more detail, beyond excluding the false positives, all the subsequent cuts are throwing away mostly valid planets and may have the risk to introduce artificial features into the sample. For example, the cut at 100-day orbital period may alter the overall number ratio of small versus large exoplanets. As demonstrated by our other paper (Survival Function Analysis of Planet Orbit Distribution, submitted to ApJ, arXiv:1801:03994 (Zeng et al., 2018)), large exoplanets ($>4~R_\oplus$) are relatively depleted inside 0.4 AU compared to small exoplanets ($<4~R_\oplus$). More specifically, large exoplanets ($>4~R_\oplus$) have a different statistical distribution in their semi-major axis or period, which is uniform in the square-root of semi-major axis, compared to small exoplanets ($<4~R_\oplus$), which is uniform in the logarithm of semi-major axis, within 0.4 AU or 100-day orbit, up to the inner threshold of 0.05 AU. Therefore, making a cut at 100-day orbital period will lose quite a few of large exoplanets ($>4~R_\oplus$).

Anyway, if one applies the strictest criteria and adopts the all the cuts, shown also in **Figure 2**, the general trend of survival function is clear in the two breaks in the power-law, and the identification that the slope in between 4 and $10~R_\oplus$ to be shallower than either side, indicating a relative paucity of planets in between $4$-$10~R_\oplus$, namely, the sub-Saturnian desert there.

**Result:**

As shown in **Figure 1** and **Figure 2**, the breaks (**3.9 ± 0.1 R$_\oplus$** and **10.3 ± 0.1 R$_\oplus$** to be exact) in the survival function power-law are the natural (model-free) divides of different regimes of planetary objects, in addition to the gap in exoplanet population histogram detected around **2 R$_\oplus$** (Fulton et al., 2017; Zeng et al., 2017a, 2017b). Thus, we propose the following classification schemes based on planet radius:

- **<4 R$_\oplus$**, small planets. They can be further divided into two sub-groups: **<2 R$_\oplus$** and **2-4 R$_\oplus$**. The small planets are gas-poor, with gaseous envelope mass fraction ($f_{env}$ = $M_{env}/M_{planet}$) less than about 5~10%. The upper bound of whether ~5% or ~10% depends on the assumptions of core mass, core composition (core here refers to the solid part of the planet), envelope thermal profile and envelope metallicity. The details of calculations can be seen in (Ginzburg et al., 2017, 2016; Lopez and Fortney, 2014). If one assumes a water-rich core like that of Uranus or Neptune, then this upper bound is more like ~5%. If one assumes a rocky core, then this upper bound is more like ~10%.
- **4-10 R$_\oplus$**, transitional planets. Statistics of the *Kepler* planet candidates shows that this group of planets follows a power-law distribution as: $dN \propto R^{-\alpha} * dR$, where $\alpha \approx 2$ (1.9±0.1 to be exact). The power index $\alpha$ in this radius range is shallower than ranges above and below, which means a relative paucity of planets per logarithmic interval of radius. This is the confirmation of a <u>sub-Saturnian desert</u>. We name them "transitional planets" as they form the bridge between small exoplanets (<4 R$_\oplus$) and gas giants (>10 R$_\oplus$).
- **>10 R$_\oplus$**, gas giants. They are dominated by H$_2$-He in bulk composition and are massive. They include Jupiter-sized planets, brown dwarfs, and even small stars.

**Discussion:**

**(1) Speculation on the origin of the kink at 4 R$_\oplus$**

It is very interesting, perhaps not by coincidence, that our own solar system happens to have planets at the kinks (~4 R$_\oplus$ for Uranus and Neptune, ~10 R$_\oplus$ for Jupiter and Saturn). It is generally known that Uranus and Neptune are icy giants, that is, their interior compositions are dominated by a mixture of ices, surrounded by about a few up to 10 percent H$_2$/He envelope (Hubbard et al., 1991; Podolak et al., 2000).

(Ginzburg et al., 2016) suggests that the intrinsic luminosity coming out of the cooling cores of young planets themselves can gradually blow off small envelope ($f_{env}$ <~5%), even without the need of stellar irradiation. However, for $f_{env}$ >~5%, planets can largely keep their envelopes over billion-year timescale. This is likely the origin of the divide at **4 R$_\oplus$**. (Ginzburg et al., 2017) tries to use this argument to explain the radius gap of small exoplanets at ~2 R$_\oplus$ by convolving this bi-modal envelope mass fraction with a smaller core mass distribution around 3~5 Earth masses (M$_\oplus$) (similar to what was assumed in (Owen and Wu, 2017)). Instead this argument works better to explain this divide at **4 R$_\oplus$**, when convolved with core size of 2~2.5 R$_\oplus$, to be consistent with higher core masses (5~20 M$_\oplus$) observed from the radial-velocity measurements and considering water-rich cores. **4 R$_\oplus$** separates core-dominated planets from gaseous-envelope-dominated planets. Since the sizes of water-rich cores of 5~20 $M_\oplus$ are 2~2.5 $R_\oplus$, **4 $R_\oplus$** is roughly the divide of envelope thickness ≲ core radius and envelope thickness ≳ core radius.

It is theorized that Earth once had a few $10^3$ bar $H_2$/He envelope when the solar nebula gas disk was around, preserved as noble gas signature in present-day Earth mantle due to early exchange/in-gassing of this primordial gaseous envelope with the early Earth magma ocean (Harper and Jacobsen, 1996). If this estimate is correct, this few $10^3$ bar envelope makes up only a few thousandth of the total mass of the Earth, which is well below the $f_{env}$~5% threshold (Ginzburg et al., 2016). Thus, Earth was not able to hold on to this envelope for long duration after the gas disk was dispersed, due to a combination of spontaneous-driven and stellar-driven losses. Similar story likely applies to most rocky cores, as it is difficult for them to grow massive enough to accrete a significant gaseous envelope in the first place, and then hold on to this envelope after gas disk dispersal, if without the additional help of ices to increase their mass by two-fold or so.

**(2) Comparison to previous works**
(2.1) Host-star metallicity
The divide at **4 $R_\oplus$** confirms previous work done by (Buchhave et al., 2014) through looking at the host stars' metallicities. (Buchhave et al., 2014) shows that the host stars of planets with R>4 $R_\oplus$ are significantly metal-enriched (0.18 ± 0.02 dex), compared to the host stars of planets with R<4 $R_\oplus$ which have solar-like metallicities. Recent study by (Winn et al., 2017) of the ultra-short-period (USP, P< 1 day, R<2 $R_\oplus$) planets further supports this analysis of metallicity. (Winn et al., 2017) shows that the metallicity distributions of USP planet and hot-Jupiter hosts are significantly different, suggesting that the USP planets are not dominated by the evaporated cores of hot Jupiters. Furthermore, the metallicity distribution of stars with USP planets is indistinguishable from that of stars with short period (1-10 day) planets with sizes between 2 and 4 $R_\oplus$. Thus, **4 $R_\oplus$** can be inferred as the divide of gas-poor and gas-rich planets. Planets above 4 $R_\oplus$ have substantial gaseous envelopes and their host stars are statistically metal-enriched compared to solar metallicity: this is the starting point of the giant-planet-metallicity correlation (Fischer and Valenti, 2005; Wang and Fischer, 2015).

(2.2) Comparison to previous claimed rapid drop-offs in planet fraction at lower radii
In (Fressin et al., 2013), the radius domain of what they call "Small Neptunes" (2-4 $R_\oplus$) exactly corresponds to the boundary and identification of water worlds in our manuscript, see their Figure 7. On page 13, they also recognized that "the increase in planet occurrence towards smaller radii from these objects is very steep". In their analysis, they also attempted to place a boundary within "Small Neptunes" (2-4 $R_\oplus$) at 2.8 $R_\oplus$. But this was done in an artificial way, quotation: "we find that dividing the small Neptunes into two subclasses (two radius bins of the same logarithmic size: 2-2.8 $R_\oplus$ and 2.8-4 $R_\oplus$), we are able to obtain a much closer match to the KOI population (K-S probability of 6%) with similar logarithmic distributions within each sub-bin as assumed before". The key here is that they have <u>assumed</u> logarithmic distribution of planet sizes within each planet category. But this assumption is not valid according to what data show in this analysis. As demonstrated in our survival function analysis, the planet size distribution fits with piece-wise power-law. And in our analysis, there is no obvious change of slope occurring at 2.8 $R_\oplus$. So, the boundary at 2.8 $R_\oplus$ is simply chosen because it sits at the logarithmic mid-point between 2 and 4 $R_\oplus$.

In (Petigura et al., 2013), they chose the same radius bins as that of (Fressin et al., 2013), see their Figure 3: a histogram showing the counts of planets within each bin. As explained above, 2.8 $R_⊕$ is chosen as the logarithmic mid-point between 2 $R_⊕$ and 4 $R_⊕$. The fact that planets in between 2-2.8 $R_⊕$ are many more than 2.8-4 $R_⊕$ is simply a manifestation of the steep slope of the power-law distribution is this radius range, instead of a change of slope there. (Silburt et al., 2015) also adopt the same binning as (Fressin et al., 2013) and (Petigura et al., 2013). In all these analyses, 2.8 $R_⊕$ is chosen but not detected.

### (3) Slope of power-law distribution

The probability distribution of the transitional planets (4-10 $R_⊕$) is best fitted to a power-law: $dN ∝ R^{-α} * dR$, where the power index $α ≈ 2$ (1.9±0.1 to be exact). It likely implies a power-law distribution in planet mass as well, but this of course depends on the exact mass-radius relationship of planets in this radius range. Many natural phenomena follow a power-law distribution with power index $α$ typically in the range of 2~3 (Clauset et al., 2009; Newman, 2005). The examples include the frequency of use of words, magnitude of earthquakes, diameter of moon craters, population of US cities, etc. (Newman, 2005) There are many mechanisms proposed for generating power-law, such as via preferential attachment, multiplicative processes, random walks, phase transitions and critical phenomena, etc. (Mitzenmacher, 2003; Newman, 2005). In our case of transitional planets, the physical mechanism accounting for the power index of $α ≈ 2$ could be a combination of the behavior of the equations of states (EOS) of $H_2$/He envelope in the planet interiors and the formation and growth processes of these planets.

The SF of small exoplanets (<4 $R_⊕$) suggests an overabundance of them over the power-law SF (red line in **Figure 1**) of transitional planets (4-10 $R_⊕$). The probability distribution of small exoplanets (<4 $R_⊕$) can be fitted to two log-normal distributions corresponding to the two peaks of the bi-modal distribution (Fulton et al., 2017; Zeng et al., 2017a, 2017b), or less likely other types of distributions. The other possibilities cannot be ruled out completely due to the detection incompleteness of the *Kepler* pipelines for planets below a certain radius threshold (Burke and Catanzarite, 2017; Christiansen et al., 2016; Christiansen and L., 2017). This threshold is the transition point from very complete detection to very incomplete detection due to pipeline incompleteness, meaning that due to low signal-to-noise ratio and other factors. It can vary from 1~2 $R_⊕$ for the orbital period range concerned (Fulton et al., 2017), however, due to the way SF is defined here, it does not affect the profile of SF above this threshold, and thus, it does not affect the breaks in power-law identified in this paper. The SF beyond ~20 $R_⊕$ suffers more from Poisson fluctuation due to low count of objects at large radii.

**Summary**

In summary, the survival function analysis provides a model-independent way to assess and classify different regimes of planetary objects according to their sizes. The boundaries identified at 3.9 ± 0.1 **$R_⊕$** and 10.3 ± 0.1 **$R_⊕$** provide constraints that any model of planet formation or interior structure should satisfy. This result will be tested against the findings of the upcoming *TESS* (Transiting Exoplanet Survey Satellite) mission.

**References:**

Akeson, R.L., Chen, X., Ciardi, D., Crane, M., Good, J., Harbut, M., Jackson, E., Kane, S.R., Laity, A.C., Leifer, S., Lynn, M., Mcelroy, D.L., Papin, M., Plavchan, P., Ramírez, S. V, Rey, R., Braun, K. Von, Wittman, M., Abajian, M., Ali, B., Beichman, C., Beekley, A., Berriman, G.B., Berukoff, S., Bryden, G., Chan, B., Groom, S., Lau, C., Payne, A.N., Regelson, M., Saucedo, M., Schmitz, M., Stauffer, J., Wyatt, P., Zhang, A.A., 2013. The NASA Exoplanet Archive: Data and Tools for Exoplanet Research. Publ. Astron. Soc. Pacific 125, 989.

Buchhave, L.A., Bizzarro, M., Latham, D.W., Sasselov, D., Cochran, W.D., Endl, M., Isaacson, H., Juncher, D., Marcy, G.W., 2014. Three regimes of extrasolar planet radius inferred from host star metallicities. Nature 509, 593–595. https://doi.org/10.1038/nature13254

Burke, C.J., Catanzarite, J., 2017. Planet Detection Metrics: Per-Target Detection Contours for Data Release 25. Kepler Sci. Doc. KSCI-19111-002, Ed. by Michael R. Haas Natalie M. Batalha.

Christiansen, J.L., Clarke, B.D., Burke, C.J., Jenkins, J.M., Bryson, S.T., Coughlin, J.L., Mullally, F., Thompson, S.E., Twicken, J.D., Batalha, N.M., Haas, M.R., Catanzarite, J., Campbell, J.R., Uddin, A.K., Zamudio, K., Smith, J.C., Henze, C.E., 2016. Measuring Transit Signal Recovery in the Kepler Pipeline. III. Completeness of the Q1-Q17 DR24 Planet Candidate Catalogue with Important Caveats for Occurrence Rate Calculations. Astrophys. J. 828, 99.

Christiansen, J.L., L., J., 2017. Planet Detection Metrics: Pixel-Level Transit Injection Tests of Pipeline Detection Efficiency for Data Release 25. Kepler Sci. Doc. KSCI-19110-001, Ed. by Michael R. Haas Natalie M. Batalha.

Clauset, A., Rohilla Shalizi, C., J Newman, M.E., 2009. Power-Law Distributions in Empirical Data. Soc. Ind. Appl. Math. Rev. 661–703.

Feigelson, E.D., Nelson, P.I., 1985. Statistical methods for astronomical data with upper limits. I - Univariate distributions. Astrophys. J. 293, 192. https://doi.org/10.1086/163225

Fischer, D.A., Valenti, J., 2005. The Planet-Metallicity Correlation. ApJ 622, 1102–1117. https://doi.org/10.1086/428383

Fressin, F., Torres, G., Charbonneau, D., Bryson, S.T., Christiansen, J., Dressing, C.D., Jenkins, J.M., Walkowicz, L.M., Batalha, N.M., 2013. The False Positive Rate of Kepler and the Occurrence of Planets. Astrophys. J. 766, 81. https://doi.org/10.1088/0004-637X/766/2/81

Fulton, B.J., Petigura, E.A., Howard, A.W., Isaacson, H., Marcy, G.W., Cargile, P.A., Hebb, L., Weiss, L.M., Johnson, J.A., Morton, T.D., Sinukoff, E., Crossfield, I.J.M., Hirsch, L.A., 2017. The California-Kepler Survey. III. A Gap in the Radius Distribution of Small Planets. Astron. J. 154, 109. https://doi.org/10.3847/1538-3881/aa80eb

Ginzburg, S., Schlichting, H.E., Sari, R., 2017. Core-powered mass loss sculpts the radius distribution of small exoplanets 6, 1–6.

Ginzburg, S., Schlichting, H.E., Sari, R., 2016. Super-Earth Atmospheres: Self-Consistent Gas Accretion and Retention. ApJ 825, 29. https://doi.org/10.3847/0004-637X/825/1/29

Harper, C.L.J., Jacobsen, S.B., 1996. Noble Gases and Earth's Accretion. Science (80-. ). 273, 1814–1818.

Hubbard, W.B., Nellis, W.J., Mitchell, A.C., Holmes, N.C., McCandless, P.C., Limaye, S.S., 1991. Interior structure of Neptune - Comparison with Uranus. Science (80-. ). 253, 648. https://doi.org/10.1126/science.253.5020.648

Lopez, E.D., Fortney, J.J., 2014. Understanding the Mass-Radius Relation for Sub-neptunes: Radius as a Proxy for Composition. ApJ 792, 1. https://doi.org/10.1088/0004-637X/792/1/1

Mitzenmacher, M., 2003. A brief history of generative models for power law and lognormal distributions. Internet Math. 1, 226–251. https://doi.org/10.1080/15427951.2004.10129088

Newman, M.E.J., 2005. Power laws, Pareto distributions and Zipf's law. Contemp. Phys. 46, 323–351. https://doi.org/10.1080/00107510500052444

Owen, J.E., Wu, Y., 2017. The Evaporation Valley in the Kepler Planets. ApJ 847, 29. https://doi.org/10.3847/1538-4357/aa890a

Petigura, E.A., Howard, A.W., Marcy, G.W., 2013. Prevalence of Earth-size planets orbiting Sun-like stars. Proc. Natl. Acad. Sci. U. S. A. 110, 19273–8. https://doi.org/10.1073/pnas.1319909110

Podolak, M., Podolak, J.I., Marley, M.S., 2000. Further investigations of random models of Uranus and Neptune. Planet. Space Sci. 48, 143–151. https://doi.org/10.1016/S0032-0633(99)00088-4

Silburt, A., Gaidos, E., Wu, Y., 2015. A STATISTICAL RECONSTRUCTION OF THE PLANET POPULATION AROUND *KEPLER* SOLAR-TYPE STARS. Astrophys. J. 799, 180. https://doi.org/10.1088/0004-637X/799/2/180

Virkar, Y., Clauset, A., 2014. Power-law distributions in binned empirical data. Ann. Appl. Stat. 8, 89–119. https://doi.org/10.1214/13-AOAS710

Wang, J., Fischer, D.A., 2015. Revealing a universal planet-metallicity correlation for planets of different solar-type stars. Astron. J. 149, 14. https://doi.org/10.1088/0004-6256/149/1/14

Winn, J.N., Sanchis-Ojeda, R., Rogers, L., Petigura, E.A., Howard, A.W., Isaacson, H., Marcy, G.W., Schlaufman, K., Cargile, P., Hebb, L., 2017. Absence of a metallicity effect for ultra-short-period planets. ApJ 154, 60. https://doi.org/10.3847/1538-3881/aa7b7c

Zeng, L., Jacobsen, S.B., Hyung, E., Vanderburg, A., Lopez-Morales, M., Sasselov, D.D., Perez-Mercader, J., Petaev, M.I., Latham, D.W., Haywood, R.D., Mattson, T.K.R., 2017a. Planet size distribution from the Kepler mission and its implications for planet formation [WWW Document]. Lunar Planet. Sci. Conf. URL http://adsabs.harvard.edu/abs/2017LPI....48.1576Z

Zeng, L., Jacobsen, S.B., Sasselov, D.D., 2017b. Exoplanet Radius Gap Dependence on Host Star Type. RNAAS.

Zeng, L., Jacobsen, S.B., Sasselov, D.D., Vanderburg, A., 2018. Survival Function Analysis of Planet Orbital Period and Semi-major Axis Distribution. Submitt. to ApJ.

**Acknowledgements**: This work was partly supported by a grant from the Simons Foundation (SCOL [award #337090] to L.Z.). Part of this research was also conducted under the Sandia Z Fundamental Science Program and supported by the Department of Energy National Nuclear Security Administration under Award Numbers DE-NA0001804 and DE-NA0002937 to S. B. Jacobsen (PI) with Harvard University. This research is the authors' views and not those of the DOE. Sandia National Laboratories is a multimission laboratory managed and operated by National Technology and Engineering Solutions of Sandia, LLC., a wholly owned subsidiary of Honeywell International, Inc., for the U.S. Department of Energy's National Nuclear Security Administration under contract DE-NA-0003525.

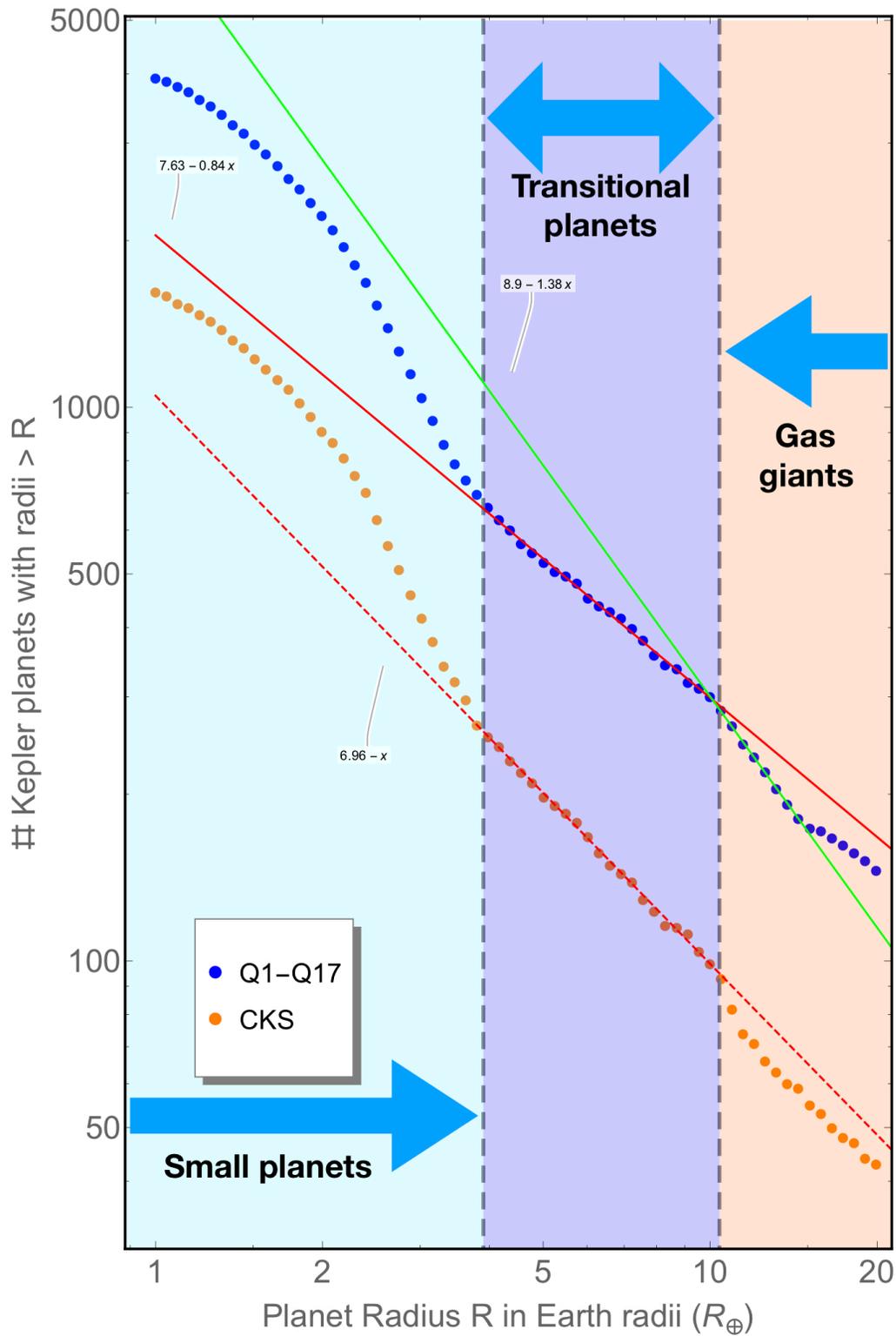

**Figure 1.** Survival function of planet radius of the *Kepler* planet candidates (4433 from Q1-Q17 Data Release 25 of *NASA* Exoplanet Archive (Akeson et al., 2013), versus 1861 from California-Kepler Survey (Fulton et al., 2017) with improved stellar parameters, both with false positives excluded already). The fits to power-law are in natural-logarithm and the proposed divides of different planet regimes are shown as vertical dashed lines.

Survival Function Analysis of Planet Size Distribution: Zeng, Jacobsen, Sasselov, Vanderburg

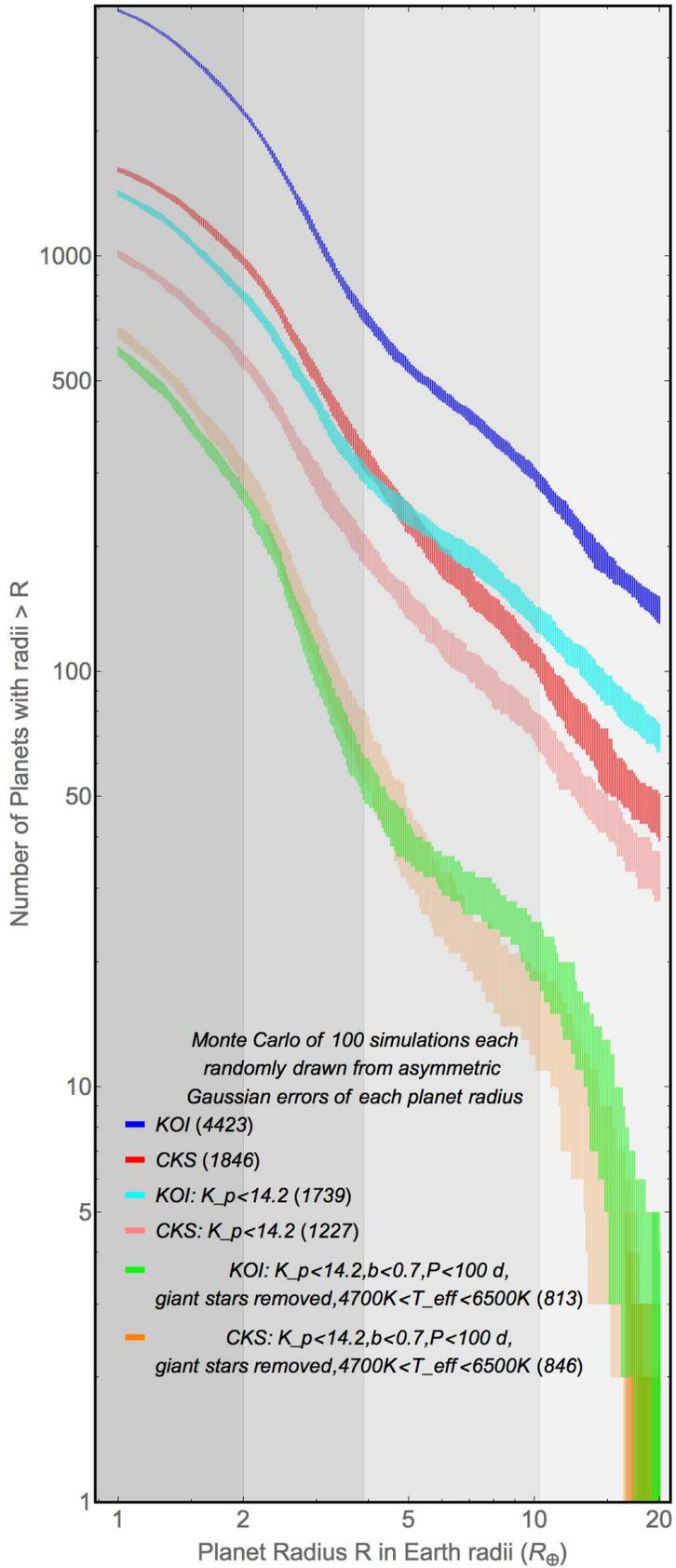

**Figure 2**. Survival function after applying successive cuts in conformity with the (Fulton et al., 2017) and taking into account error in planet radius by Monte-Carlo method by drawing randomly from the Gaussian with asymmetric uncertainty on each side, centered at the best-fit value of radius of each planet, and plot the survival function. We repeat the whole process 100 times, and calculate 100 survival functions for each scenario, to give an idea as to the uncertainty in the survival function itself.